\documentclass[a4paper,amsmath,amssymb]{revtex4}
\pdfoutput=1
\usepackage{geometry}                
\usepackage{graphicx}
\usepackage{amssymb}
\usepackage{epstopdf}
\usepackage{float}
\usepackage{color}

\newcommand{\be}{\begin{eqnarray}} 
\newcommand{\ee}{\end{eqnarray}}
 
\newcommand{\D}{\mathrm{d}}

\date{\today}
                              
\begin{document}
\title{Calculation of the Voronoi boundary for lens-shaped particles and spherocylinders}

\author{Louis Portal$^{1}$, Maximilien Danisch$^{1,3}$, Adrian Baule$^{1,2}$, Romain Mari$^{1}$ \& Hern\'an A. Makse$^{1}\footnote{Correspondence to: hmakse@lev.ccny.cuny.edu}$}

\affiliation{ $^1$Levich Institute and Physics Department, City
  College of New York, New York, New York 10031, USA \\ $^2$School of
  Mathematical Sciences, Queen Mary University of London, London E1
  4NS, UK\\
  $^{3}$Laboratoire d'Informatique de Paris 6, Universit\'e Pierre et Marie Curie, 4 Place Jussieu, 75005 Paris, France
  }

\begin{abstract}
We have recently developed a mean-field theory to estimate the packing fraction of non-spherical particles [A. Baule {\it et al}, Nature Commun. (2013)]. The central quantity in this framework is the Voronoi excluded volume, which generalizes the standard hard-core excluded volume appearing in Onsager's theory. The Voronoi excluded volume is defined from an exclusion condition for the Voronoi boundary between two particles, which is usually not tractable analytically. Here, we show how the technical difficulties in calculating the Voronoi boundary can be overcome for lens-shaped particles and spherocylinders, two standard prolate and oblate shapes with rotational symmetry. By decomposing these shapes into unions and intersections of spheres analytical expressions can be obtained.
\end{abstract}
  
\maketitle

\section{Introduction}

Packings of hard particles are ubiquitous in many fields in science and engineering \cite{Torquato10}. Most studies have focused on particles with spherical symmetry, which in a disordered arrangement typically achieve packing fractions of $\approx 64\%$ volume fraction. By contrast, both prolate and oblate non-spherical shapes can achieve higher packing fractions, as found in simulations \cite{Williams03,Abreu03,Donev04} and experiments \cite{Man05}. The existence of strong positional and orientational correlations has so far prevented any systematic study apart from the limit of infinitely thin rods, which are treated in an equilibrium setting by Onsager's virial expansion of the free energy \cite{Onsager49}. However, the densest packings are typically found in a regime close to the sphere \cite{Donev04}, for which this expansion breaks down.

A statistical mechanical framework to treat jammed granular matter has been proposed more than two decades ago by S.~F. Edwards, who postulated that the macroscopic properties of these systems can be calculated as ensemble averages similar to equilibrium systems \cite{Edwards89}. These averages are taken over all jammed microstates at a fixed system volume, where all microstates are assumed as equiprobable. This means that the role of energy in thermal systems is replaced by the volume in granular systems, leading to an analogous statistical mechanical framework. The main task is then to obtain the volume function (analogue of the Hamiltonian), which describes the system volume as a function of the particles' positions and orientations. Here, different conventions can be employed to partition the total volume into cells associated with each particle \cite{Ball02,Blumenfeld03}, the simplest of which is the Voronoi tesselation \cite{Okabe,Makse04}. However, in 3D these exact volume functions are difficult to handle analytically, requiring a suitable coarse-graining procedure.

We have recently followed such a mesoscopic approach in order to develop a mean-field theory of packings of both spheres and anisotropic particles \cite{Song08,Song10,Baule13}. The central quantity in our framework is the coarse-grained Voronoi volume $W(z)$ of a single particle, which, roughly speaking, contains the volume that is closer to this particle than to any other one on average. This approach is further motivated by the observation that, as the particle number $N\to\infty$, packings show reproducible phase behaviour and are characterized by only few observables such as the packing fraction and the average coordination number $z$ \cite{Makse04,Brujic05,Brujic07}. $W(z)$ satisfies a self-consistent equation: \cite{Song08,Song10,Baule13} 
\begin{equation}
\label{self}
W(z)=\int\D\mathbf{c}\exp\left\{-\frac{V^*(\mathbf{c})}{W-V_0}-\sigma(z)S^*(\mathbf{c})\right\},
\end{equation}
where the key ingredients $V^*(\mathbf{c})$ and $S^*(\mathbf{c})$ are referred to as the Voronoi excluded volume and surface, which extend Onsager's hard-core excluded volume to jammed packings. In Eq.~(\ref{self}), the quantity $\sigma(z)$ is the free surface density for a packing with $z$ contacts and is obtained from simulations of local configurations. Once $W(z)$ is determined by solving the self-consistency equation, the packing fraction as a function of $z$ follows simply from $\phi=V_0/W(z)$, where $V_0$ is the particle volume. This approach has to be complemented with a prescription for the value of $z$ in the packing, which is fixed by the isostatic conjecture $z=2d_{\rm f}$ satisfied by spheres. However, anisotropic particles can have $z< 2d_{\rm f}$ due to redundancy in the force and torque balance equations that define mechanical equilibrium. A quantitative theory for this effect has been developed in Ref.~\cite{Baule13} and calculates $z$ as an average $z=2\left<d_{\rm eff}\right>$ over effective number of degrees of freedom. 

Therefore, in order to apply our framework the main task is to calculate $V^*(\mathbf{c})$ and $S^*(\mathbf{c})$ for a particular particle shape. Both quantities are defined from an exclusion condition on the Voronoi boundary (VB) between two particles, so that the calculation requires analytic expressions for the VB, which are typically difficult to obtain \cite{Okabe,Boissonnat06}. In this paper we show how to calculate the VB of lens-shaped particles and spherocylinders, two model shapes for both oblate and prolate anisotropic particles. Regular crystal packings of these shapes have recently been investigated theoretically in Ref.~\cite{Torquato12}. For both shapes analytic expressions for the VB can be obtained, which is in contrast to, e.g., prolate and oblate ellipsoids. The underlying reason is that both can be decomposed into decompositions and intersections of spheres of equal radii, such that the VB is generated from simpler effective interactions, namely those between points, lines, and anti-points leading to an exact algorithm for the VB \cite{Baule13}. By comparison, the corresponding decomposition of prolate and oblate ellipsoids requires a dense set of spheres with continuously varying radii, which greatly complicates the problem. In the following we explicitly show how to calculate the VB for lens-shaped particles and spherocylinders following our algorithm. This will guide the calculation of the VB between more complicated shapes that can be decomposed similarly.

This paper is organized as follows. In the next section we first recapitulate how to calculate the VB between spheres before showing how to extend it to lens-shaped particles (Sec.~\ref{Sec_lensvb}) and how to calculate the contact radius between two such objects (Sec.~\ref{Sec_lensrad}). Then we review the calculation of both quantities for spherocylinders (Sec.~\ref{Sec_spherovb} and Sec.~\ref{Sec_spherorad}). We summarize how to use these results in order to evaluate the excluded volume and surface of these particles in Sec.~\ref{Sec_exvols}. We finally conclude with a brief discussion of potential generalizations (Sec.~\ref{Sec_discussion}).

\section{Results}

We set the centre of our coordinate system to the centre of mass of particle $i$ and fix the orientation of this particle along $\mathbf{\mathbf{\hat{z}}}$. Given a direction $\mathbf{\hat{c}}$, a point on the VB is found at $s\mathbf{\hat{c}}$, where $s$ depends on the position $\mathbf{r}$ and orientation $\mathbf{\hat{t}}$ of particle $j$: $s=s(\mathbf{r},\mathbf{\hat{t}},\mathbf{\hat{c}})$. The value of $s$ is obtained from two conditions:
\begin{enumerate}
\item The point $s\mathbf{\hat{c}}$ has the minimal distance to each of the two objects along the direction $\mathbf{\hat{c}}$.
\item Both distances are the same.
\end{enumerate}
The VB between two spheres of equal radii is the same as the VB between two points at the centres of the spheres. Therefore, condition $1$ is trivially satisfied for every $s$ and condition $2$ translates into the equation
\be
\label{VBspherecond}
(s\mathbf{\hat{c}})^2=(s\mathbf{\hat{c}}-\mathbf{r})^2,
\ee
leading to
\be
\label{VBsphere}
s=\frac{r}{2\mathbf{\hat{c}}\mathbf{\hat{r}}},
\ee
i.e., the VB is the plane perpendicular to the separation vector $\mathbf{r}$ at half the separation. Already for two spheres of unequal radii, the VB is a curved surface. Taking into account the different radii $a_i$ and $a_j$, Eq.~(\ref{VBspherecond}) becomes \cite{Danisch10}
\be
s-a_i=\sqrt{(s\mathbf{\hat{c}}-\mathbf{r})^2}-a_j,
\ee
which has the solution
\be
s=\frac{1}{2}\frac{r^2-(a_i-a_j)^2}{\mathbf{\hat{c}}\mathbf{\hat{r}}-(a_i-a_j)}.
\ee
Finding a solution for both conditions for general non-spherical objects is in general non-trivial. Next we show how to overcome the difficulties for lens-shaped particles.

\subsection{The VB between two lens-shaped particles}

\label{Sec_lensvb}

\begin{figure}[]
 \centering
 	\includegraphics[width=0.6\textwidth]{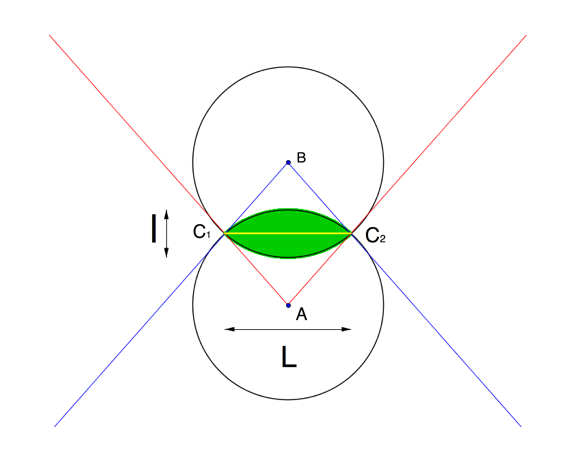}
	\caption{A lens is made of two spheres centred in A and B. The lens is the green shape, the intersection of the spheres. The yellow line denotes the ``crown'', whose extremities in the plane are marked C1 and C2.
 	}
	\label{LENS}
\end{figure}

\begin{figure}[]
 \centering
 	\fbox{\includegraphics[width=0.7\textwidth]{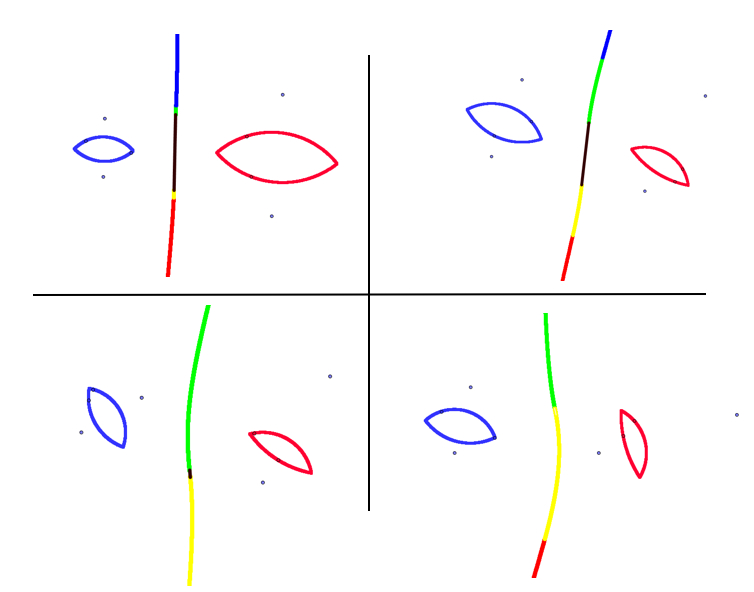}}
	
	\caption{VB of two lenses in the plane, coloured in function of 
	the type of interaction.
	Sphere-Sphere interaction: red and blue lines; 
	Sphere-Crown interaction: green and yellow lines; 
	Crown-Crown: black line.
	The four points are the positions of the centres of the spheres constituting the lenses. 
	}
	\label{Voro2D}
\end{figure}

A lens-shaped particle is made of two spherical parts and their circular junction that we call the ``crown''. Here we use two spheres of the same radius $R$ so that the gravity centre is the centre of the crown. $L$ denotes its diameter of and $l$ the thickness of the lens. The aspect ratio of a lens is then defined by $\alpha = l/L$ (see Fig.~\ref{LENS}). We first wrote a short code (using Geogebra) to visualize qualitatively the VB between two lens-shaped particles in 2D for the most general case of an intersection of two spheres with different radii (Fig.~\ref{Voro2D}).

\begin{figure}[]
 \begin{center}
 	\includegraphics[width=0.6\textwidth]{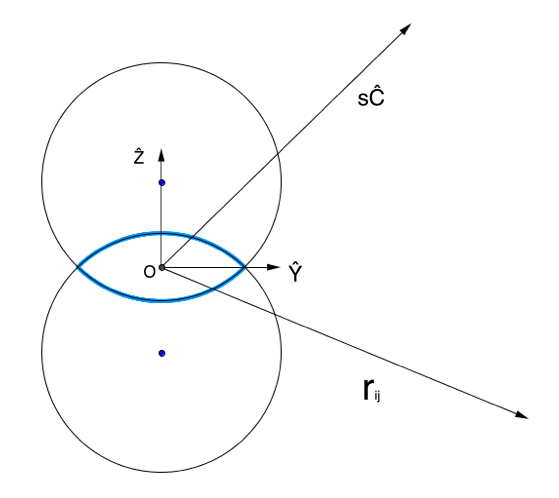}
 	\caption{Parametrization of lens $i$, blue points denote the centres of the spheres constituting the lens. 
	}
	\label{PARA}
\end{center}
\end{figure}

\begin{figure}[]
 \centering
 	\fbox{\includegraphics[width=0.4\textwidth]{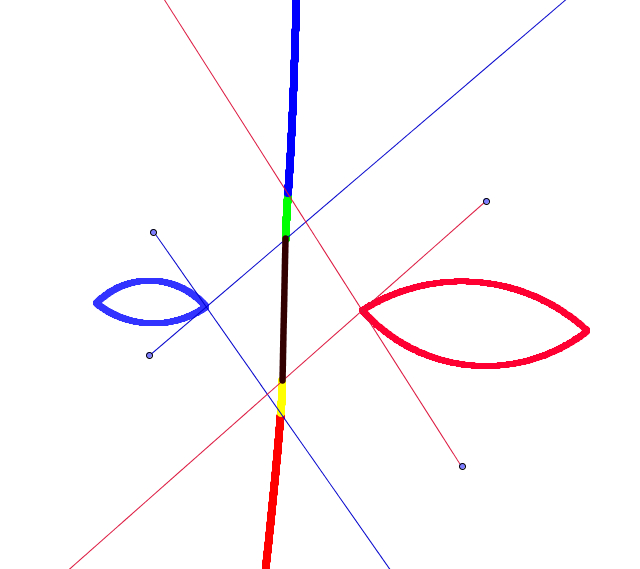}}
	\caption{Different interactions that generate the VB depending on the cone configurations}
	\label{VBCONES}
\end{figure}

We then attempted to follow the calculation explained above for spheres, with the same set of conditions to determine the VB. However, the spherical parts of the lens determine the VB only in specific regions. These are defined by cones as displayed in Fig.~\ref{LENS}. When the VB is found out of these cones, we should consider the extremity of the lens that corresponds to the circle defined by all points belonging to both spheres (the ``crown'') for the calculation. Indeed, a sphere is assimilated to its centre for the calculation because the centre, the point of the sphere that interacts and the VB are aligned. It is not valid anymore when being out of the cones for lenses. Thus if we fix two lenses in space, the VB along a given direction depends on these cones, and we have the following different types of interaction (Fig.~\ref{VBCONES}): 

\begin{enumerate}
\item{{\it Sphere-Sphere} (4 interactions): The boundary falls in a cone of each lenses. We have the same calculation as in the sphere study.}
\item{{\it Sphere-Crown} and {\it Crown-Sphere} (4 interactions): The boundary falls in a cone of the first lens but does not fall in a cone of the second. The distances to equalize are between the surface of the interacting sphere and a precise point of the interacting crown.}
\item{{\it Crown-Crown} (1 interaction): The boundary falls out of both cones for both lenses. The distances to equalize are between two precise points on the crowns.}
\end{enumerate}

The purpose of the following algorithm is to calculate each interaction one after the other and to evaluate if the VB effectively falls in the corresponding area and if it is the minimum of all interactions. If it does, then it is a valid solution. Some tests may allow one to avoid the evaluation of one or two cases.

\subsubsection{Parametrization of the problem}

We use the parameters displayed in Fig.~\ref{PARA}. The origin of the coordinate system is set to the gravity centre of lens $i$. $r\mathbf{\hat{r}}_{ij}$ links the centres of lens $i$ and $j$. The VB is to be found along vector $\mathbf{\hat{c}}$. We can set $\mathbf{\hat{c}}$ in the plane $(\mathbf{\hat{y}},\mathbf{\hat{z}})$ because of the axisymmetry of the lens. The orientation of lens $j$ in the space is given by the vector $\mathbf{\hat{z}}_j$. We are interested in the calculation of $s$ along $\mathbf{\hat{c}}$, where $s$ denotes the position of the VB between lens $i$ and lens $j$. As mentioned before we have the same set of conditions as in the sphere study:

\begin{enumerate}
\item The point $s\mathbf{\hat{c}}$ has the minimal distance to each of the two objects along the direction $\mathbf{\hat{c}}$.
\item Both distances are the same.
\end{enumerate}

Depending on the direction of $\mathbf{\hat{c}}$, $s$ may necessarily be in a cone of i, or not. It is determined by an angular test.
If the direction of $\mathbf{\hat{c}}$ stays in a cone of lens $i$, the spherical part of lens $i$ whose centre is the vertex of this cone, will interact. It is assimilated to the centre of this sphere. Thus, the square of the distance between $s\mathbf{\hat{c}}$ and the concerned sphere $i$ is: 
\be
\label{DeiS}
D_i^2 &=& (\pm(R-\frac{l}{2})\mathbf{\hat{z}} + s\mathbf{\hat{c}})^2.
\ee
If $s$ may be found outside of both cones of lens $i$, depending of the orientation of $\mathbf{\hat{c}}$ relatively to the cones of lens $i$, one should also evaluate a crown interaction for $i$. In this case, the projection of $\mathbf{\hat{c}}$ on the crown $i$ will interact since it is the closest point of lens $i$ to $s$, changing Eq.~($\ref{DeiS}$) into: 
\be
\label{DeiC}
D_i^2 &=& (-\frac{L}{2}\mathbf{\hat{y}} + s\mathbf{\hat{c}})^2,
\ee
We then compute all interactions with lens $j$ and keep the minimum result that falls its good corresponding area. This is the purpose of the algorithm: We try to determine the interacting part of $i$ at the beginning with an angular test and calculate each combination with $j$. For each interaction we determine if $s$ falls inside the corresponding area of influence of $j$ and is the minimum in comparison with the other cases. If it does then it is a valid solution, if not we calculate another interaction. Tests are based on an angular verification of the positions of $s$ relatively to the cones of $j$, as well as we had for $i$. Thus, if $s$ falls in a cone of $j$ 
\be
\label{DejS}
D_j^2 &=& (\pm(R-\frac{l}{2})\mathbf{\hat{z}}_j -r\mathbf{\hat{r}}_{ij}+s\mathbf{\hat{c}})^2.
\ee
If $s$ falls outside of both cones $j$, the point of crown $j$ that interacts is necessarily the one defined by the projection of the vector $-r\mathbf{\hat{r}}_{ij}+s\mathbf{\hat{c}}$ on the crown of lens $j$. It is given by $\mathbf{t_j}$: 
\be
\mathbf{t}_j = \frac{L}{2N} ({\mathbf{\hat{z}}_j}\times(( -r\mathbf{\hat{r}}_{ij}+s\mathbf{\hat{c}})\times{\mathbf{\hat{z}}_j}))
\ee
and
\be
N^2=r^2  (\mathbf{\hat{z}}_j\times(\mathbf{\hat{r}}_{ij}\times\mathbf{\hat{z}}_j))^2
-2rs(\mathbf{\hat{z}}_j\times(\mathbf{\hat{r}}_{ij}\times\mathbf{\hat{z}}_j))(\mathbf{\hat{z}}_j\times(\mathbf{\hat{c}}\times\mathbf{\hat{z}}_j))
+s^2  (\mathbf{\hat{z}}_j\times(\mathbf{\hat{c}}\times\mathbf{\hat{z}}_j))^2
\ee
Thus, 
\be
\label{DejC}
D_j^2 &=& (-\mathbf{t}_j -r\mathbf{\hat{r}}_{ij}+s\mathbf{\hat{c}})^2,
\ee
Condition $2$ requires $D_i^2=D_j^2$. However, if a sphere is assimilated to its centre, a crown is not. Indeed if we assimilate the crown to its centre, we may have solutions on the sphere centred on the crown that are not on the plane of the crown. Thus we will have to subtract the radius of a sphere in one term of the equation $D_i^2=D_j^2$ in the Sphere-Crown and Crown-Sphere interactions.

Condition $1$ is always satisfied in each calculation because we work with spheres of the same radius, and because the interacting point of the crown is necessarily defined by the projections of the vectors calculated above. Now we show the results for the different interactions:

\subsubsection{Sphere-Sphere interaction}

This case arises if $s$ falls in a cone of lens $i$ and a cone of lens $j$. $D_i^2=D_j^2$ leads to:
\be
(\pm(R-\frac{l}{2})\mathbf{\hat{z}}+s\mathbf{\hat{c}})^2=(\pm(R-\frac{l}{2})\mathbf{\hat{z}}_j-r\mathbf{\hat{r}}_{ij}+s\mathbf{\hat{c}})^2
\ee
Thus we obtain for $s$:
\be
s=\frac{1}{2}\frac{r^2\pm2(R-l/2)r\mathbf{\hat{z}}_j\mathbf{\hat{r}}_{ij}}{\pm(R-l/2)\mathbf{\hat{z}}\mathbf{\hat{c}}\pm(R-l/2)\mathbf{\hat{z}}_j\mathbf{\hat{c}}+r\mathbf{\hat{r}}_{ij}\mathbf{\hat{c}}}
\ee

\subsubsection{Crown-Sphere interaction}

In this case $s$ is outside of both cones of lens $i$. Thus, the distances to equalize are not between the 
centres of two spheres $S_i-S_j$ and $s$ but between the crown of lens $i$ and the spherical 
surface of lens $j$. The point of crown $i$ that interacts is necessarily the one defined by the 
projection of $\mathbf{\hat{c}}$ on the crown of lens $i$, and we have to subtract the radius of the concerned 
sphere $j$ ($\pm$) in the equation. Since all spheres do have the same radius we have:  
\be
(-\frac{L}{2}\mathbf{\hat{y}}+s\mathbf{\hat{c}})^2+2R\sqrt{(-\frac{L}{2}\mathbf{\hat{y}}+s\mathbf{\hat{c}})^2}+R^2=
(\pm(R-\frac{l}{2})\mathbf{\hat{z}}_j-r\mathbf{\hat{r}}_{ij}+s\mathbf{\hat{c}})^2
\ee
We then have an equation that is to be squared in order to obtain a polynomial of order 2 on $s$ : 
\begin{eqnarray}
r^2+2r(\pm(R-\frac{l}{2})\mathbf{\hat{z}}_j\mathbf{\hat{r}}_{ij}-s\mathbf{\hat{r}}_{ij}\mathbf{\hat{c}})
\pm2s(R-\frac{l}{2})\mathbf{\hat{z}}_j\mathbf{\hat{c}}&\nonumber\\
+sL\mathbf{\hat{y}}\mathbf{\hat{c}}+(R-\frac{l}{2})^2-\frac{L^2}{4}-R^2=2R\sqrt{(-\frac{L}{2}\mathbf{\hat{y}}+s\mathbf{\hat{c}})^2}
\end{eqnarray}

\subsubsection{Sphere-Crown interaction}

In this case, angular tests for the previous interactions show that $s$ does not fall inside any of both cones of lens $j$. As in the Crown-Sphere interaction we subtract the radius of a sphere in one term of the equation:
\be
(\sqrt{(\pm(R-\frac{l}{2})\mathbf{\hat{z}}+s\mathbf{\hat{c}})^2 }-R)^2=(-\mathbf{t}_j-r\mathbf{\hat{r}}_{ij}+s\mathbf{\hat{c}})^2
\label{SC}
\ee
Since the denominator $N$ of $\mathbf{t_j}$ depends on the square root of terms in $s^2, s,$ 
and constants, we square the previous equation to have a polynomial in $s$. It is a 
polynomial of order 6 that we resolve numerically.

\subsubsection{Crown-Crown interaction}

Now all the angular tests show that $s$ is out of both cones of lens $i$ and both cones of lens $j$. $D_i^2=D_j^2$ leads to: 
\be
(-\frac{L}{2}\mathbf{\hat{y}}+s\mathbf{\hat{c}})^2=(-\mathbf{t}_j-r\mathbf{\hat{r}}_{ij}+s\mathbf{\hat{c}})^2
\ee
We proceed as in Eq.~(\ref{SC}) and obtain a polynomial of order 6 that we also resolve numerically.

\subsection{Calculation of the contact radius of two lens-shaped particles}

\label{Sec_lensrad}

Still using the same parametrization, we now want to determinate the position $r^*$ of the centre of lens $j$ 
so that it is in contact with lens $i$ along $\mathbf{\hat{r}}_{ij}$. So we just consider the vector $\mathbf{\hat{r}}_{ij}$, and $\mathbf{\hat{c}}$ has no influence in this calculus. We also have a test procedure and a choice to make for $r^*$.

It consists of calculating first a Sphere-Sphere contact and then determinate if the point of contact belongs to both lens. Indeed, it may happen that the point of contact is on a sphere constituting a lens, but not on the lens itself. This determines an upper limit to $r^*$. If the test is positive, it is a valid solution. If the test is negative, then one of the crowns necessarily is in contact. 

Thus we calculate a Sphere-Crown contact and Crown-Sphere one. For each of these two new interactions we test if the point of contact belongs to both lenses. If the test is positive for both interactions the solution is valid and the same for both cases. If it is positive for only one interaction then it is a valid one. If it is negative for both interactions $r^*$ is necessarily determined by a Crown-Crown contact. It may happen since the tests are exclusive and because the resolution of the equations is numerical for the Sphere-Crown and Crown-Sphere interactions. We may also have special cases where lenses $i$ and $j$ have the same orientation with their crown in the same plane, thus a Crown-Crown interaction is directly calculated. We note that there is one and only one Sphere-Sphere interaction to evaluate: The sphere from which $\mathbf{\hat{r}}_{ij}$ goes out for $i$ and the one that it enters for $j$.

\subsubsection{Sphere-Sphere}

We directly obtain 
\be
(\pm(R-\frac{l}{2})\mathbf{\hat{z}}+r^*\mathbf{\hat{r}}_{ij}\pm(R-\frac{l}{2})\mathbf{\hat{z}}_j)^2=4R^2
\ee
The signs ($\pm$) depend on which sphere interacts, and after developing we keep the positive root of the polynomial:
\be
(r^*)^2+
2r^*(R-l/2)(\pm\mathbf{\hat{z}}\pm\mathbf{\hat{z}}_j)\mathbf{\hat{r}}_{ij}+
2(R-\frac{l}{2})^2(1\pm\mathbf{\hat{z}}\mathbf{\hat{z}}_j)-4R^2
\ee

\subsubsection{Crown-Sphere and Sphere-Crown}

These interactions are of course similar. We show here the Crown-Sphere interaction, and we follow the same procedure in a coordinate system fixed to lens $j$ to calculate the Sphere-Crown interaction. We denote by $(x,y,z)$ the point of contact. It belongs to the crown $i$, which simply means: 
\[
(S)\begin{cases} 
x^2+y^2= L^2/4\\
z=0
\end{cases}
\]
It also belongs to the projection of the sphere $j$ on the plane of crown $i$, thus we add to the previous system:  
\be
(r^*\mathbf{\hat{r}}_{ij}\mathbf{\hat{x}}\pm(R-\frac{l}{2})\mathbf{\hat{z}}_j\mathbf{\hat{x}}-x)^2+
(r^*\mathbf{\hat{r}}_{ij}\mathbf{\hat{y}}\pm(R-\frac{l}{2})\mathbf{\hat{z}}_j\mathbf{\hat{y}}-y)^2=R^2
\label{SUB}
\ee
Then we substitute $x$ from ($S$) in Eq.~(\ref{SUB}), thus : 
\be
(r^*\mathbf{\hat{r}}_{ij}\mathbf{\hat{x}}\pm(R-\frac{l}{2})\mathbf{\hat{z}}_j\mathbf{\hat{x}}-
\sqrt{\frac{L^2}{4}-y^2})^2+
(r^*\mathbf{\hat{r}}_{ij}\mathbf{\hat{y}}\pm(R-\frac{l}{2})\mathbf{\hat{z}}_j\mathbf{\hat{y}}-y)^2=R^2
\ee
We develop and simplify the expression, which gives a polynomial of order 2 in $y$. The equations and the algorithm take into account that $y$ and $r^*$ must be positive since the signs in Eq.~(\ref{SUB}) depends on the interacting sphere ($\pm(R-l/2)\mathbf{\hat{z}}_j\mathbf{\hat{y}}$ and $\pm(R-l/2)\mathbf{\hat{z}}_j\mathbf{\hat{x}}$ terms).

Thus we do necessarily have one and only one solution for $y$ so that $\Delta=0$ for the polynomial of order 2 on $y$. This leads to a polynomial of order 6 in $r^*$ that we resolve numerically. Thus we obtain $y$ and an angular test allows one to obtain $x$: 
\be
x=\pm\sqrt{L^2/4-y^2}
\ee
These coordinates $x$ and $y$ allow one to find now if the interaction is the good one with another angular test. In order to calculate the Sphere-Crown interaction we temporarily change the base that we set in lens $j$ and do the same calculation in the plane of Crown $j$.

\subsubsection{Crown-Crown}

We denote by $(x,y,z)$ the point of contact. It belongs to the crown $i$ :
\[
(S1)\begin{cases} 
x^2+y^2= L^2/4\\
z=0
\end{cases}
\]
It also belongs to the crown $j$ :
\[
(S2) \begin{cases}
(r^*\mathbf{\hat{r}}_{ij}\mathbf{\hat{x}}-x)^2+ (r^*\mathbf{\hat{r}}_{ij}\mathbf{\hat{y}}-y)^2+
r^{*2}\mathbf{\hat{r}}_{ij}\mathbf{\hat{z}}=L^2/4\\&\nonumber
\\
-x\mathbf{\hat{x}}\mathbf{\hat{z}}_j-y\mathbf{\hat{y}}\mathbf{\hat{z}}_j+r^*\mathbf{\hat{r}}_{ij}\mathbf{\hat{z}}_j=0
\end{cases}
\]
We substitute $x$ in the first equation of $(S2)$ with its expression from the second equation of $(S2)$ to have an expression on $y$. Then we follow the exact same procedure as mentioned in the Sphere-Crown interaction.

\subsection{The VB between two spherocylinders}

\label{Sec_spherovb}

\begin{figure}[]
    \centering
        \includegraphics[width=0.5\textwidth]{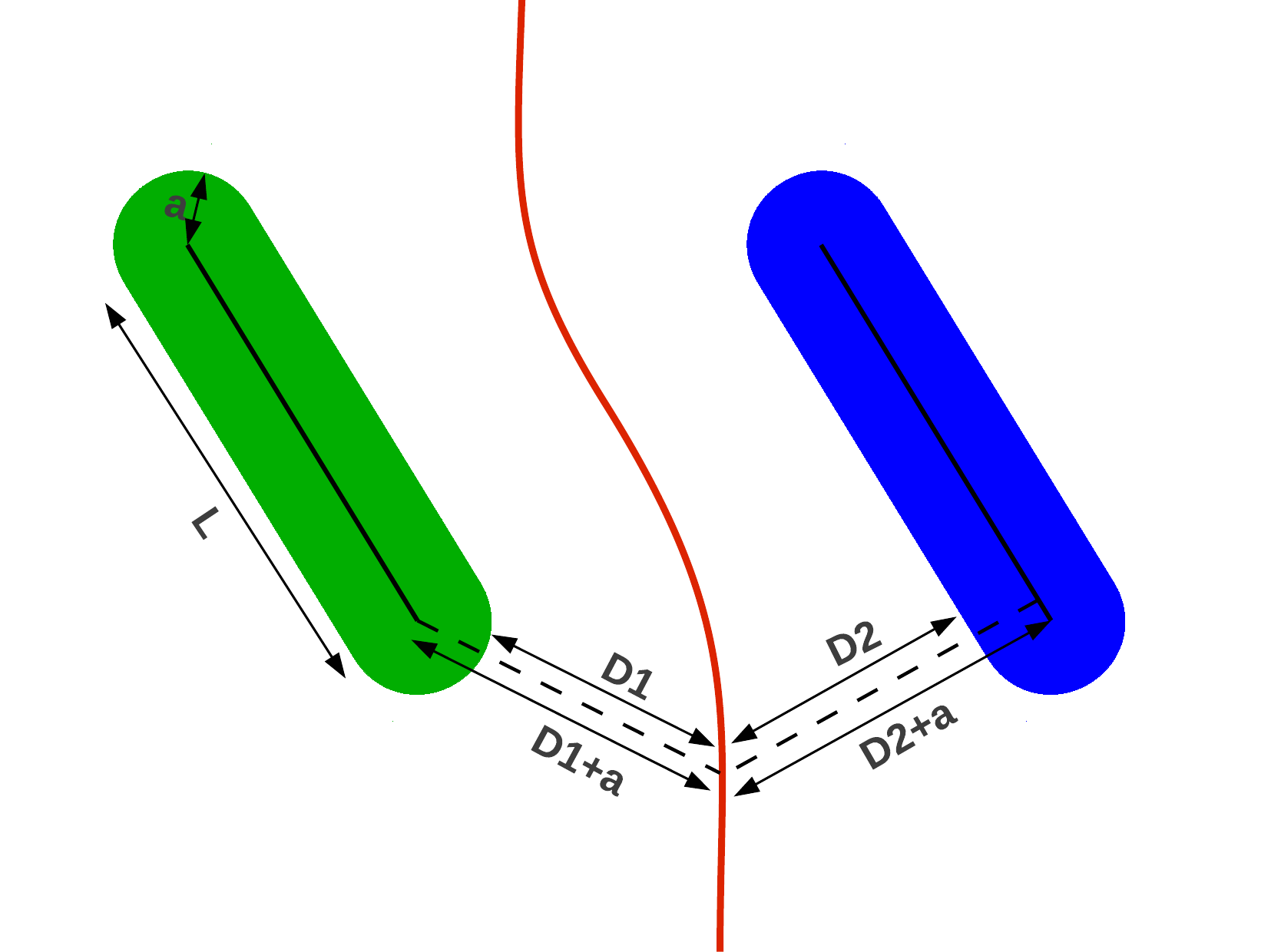}
        \caption{Voronoi diagram equivalence between two
          spherocylinders and two line segments defining the
          spherocylinders. The VB is equivalent to that
          of two rods of vanishing width and length $L$, i.e., solving
          for the point where $D_1=D_2$ is the same as solving
          $D_1+a=D_2+a$.}
    \label{rod}
\end{figure}

The calculation of the Voronoi diagram of spherocylinders is comparatively simpler than the one for, e.g., ellipsoids thanks to the following property: The VB of two spherocylinders of length $L$ (of the cylindrical part) and radius $a$ (of the semi-spherical end-caps) is equivalent to the VB between two line segments of length $L$ (see Fig.~\ref{rod}) at the centre of the cylindrical part. In the following we refer to these line segments as ``rods". This equivalence is analogous to the sphere--point equivalence of equal size spheres.

The radius $a$ of the spherocylinders thus does not appear explicitly
in the calculation of the VB as outlined in the next
sections. However, the radius enters naturally as a limiting
condition for the possible configurations of the spherocylinders and defines the contact radius $r^*(\hat{r},\hat{t})$, which is discussed in more detail in Sec.~\ref{Sec_spherorad}.

The calculation of the VB and the contact radius between two spherocylinders has previously been presented in the supplementary material of Ref.~\cite{Baule13}. In the following we use the convention that a vector $\mathbf{x}$ can be decomposed as $\mathbf{x} = x \mathbf{\hat{x}}$, where $x$ denotes the absolute value and $\mathbf{\hat{x}}$ the unit direction. The product of two vectors $\mathbf{x}\mathbf{y}$ denotes the scalar product $\mathbf{x} \cdot \mathbf{y} = \sum_k x_k y_k$, where the sum is over all components.

\begin{figure}[]
\begin{center}
\includegraphics[width=14cm]{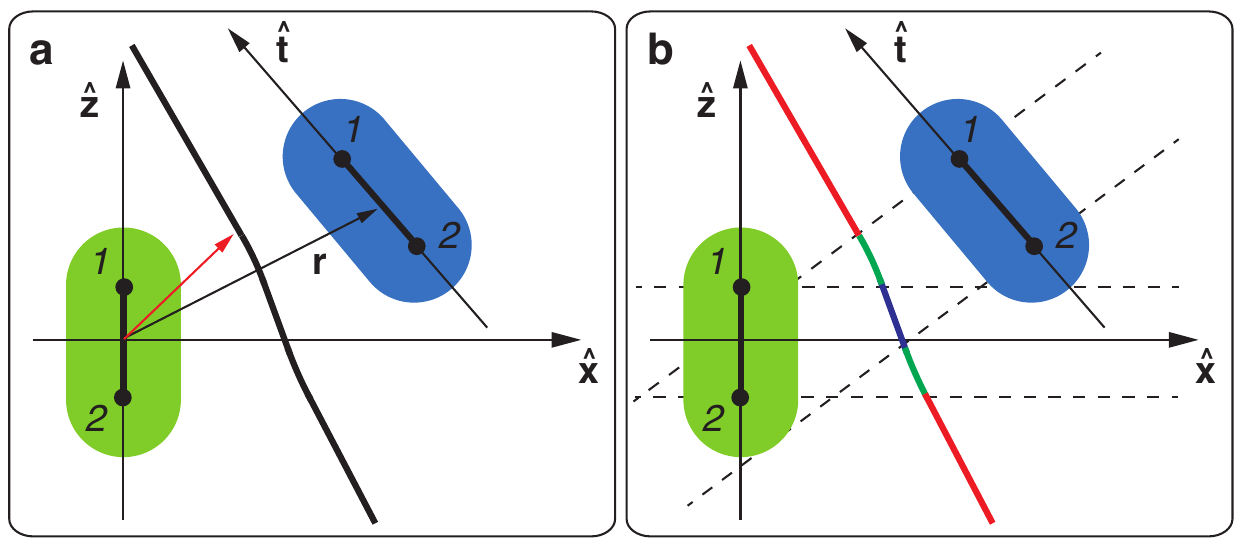} 
\caption{\label{Fig_spheros_vb} {\bf (a)} Parametrization of the VB between two spherocylinders of relative orientation $\mathbf{\hat{t}}$ and position $\mathbf{r}$. {\bf (b)} The VB consists of the VBs due to the interaction of the four points and two lines (indicated in different colours), which are separated following our algorithm. The red part of the VB is due to the Point-Point interactions, the green part due to Line-Point and Point-Line, and the blue part due to the Line-Line.}
\end{center}
\end{figure}

We align rod $i$ with the $\mathbf{\hat{z}}$ axis of our coordinate system, so that a point on it is parametrized by the vector $t_i\mathbf{\hat{z}}$ with $t_i \in [-L/2;L/2]$. Likewise, the orientation of rod $j$ is given by $\mathbf{\hat{t}}$, so that a point on rod $j$ is parameterized by $\mathbf{r}+t_j\mathbf{\hat{t}}$, where also $t_j \in [-L/2;L/2]$. See Fig.~\ref{Fig_spheros_vb}a for the setup of our coordinate ssytem. In order to solve the two conditions that define the VB, we first find the distance between $s\mathbf{\hat{c}}$ and a point on rod $i$ and $j$, denoted by $D_i^2$ and $D_j^2$, respectively. We obtain
\be
\label{di}
D_i^2 &=& (t_i\mathbf{\hat{z}} - s\mathbf{\hat{c}})^2,\\
D_j^2 &=& (\mathbf{r} + t_j\mathbf{\hat{t}} - s\mathbf{\hat{c}})^2.
\ee
Condition 1. (minimal distance) requires:
\be
\frac{\partial D_i^2}{\partial t_i}&=&0,
\label{di0}\\
\frac{\partial D_j^2}{\partial t_j}&=&0.
\label{dj0}
\label{di0dj0}
\ee
Solving these two conditions yields the minimal values
\be
\label{titj1}
t_i^{\rm min}& = &s\mathbf{\hat{c}}\mathbf{\hat{z}} = s (\mathbf{\hat{c}}\mathbf{\hat{z}}),\\
\label{titj2}
t_j^{\rm min} &=& (s\mathbf{\hat{c}}-\mathbf{r}) \mathbf{\hat{t}} = s (\mathbf{\hat{c}}\mathbf{\hat{t}})- r.
\ee
Condition 2. (equal distances) requires:
\be
D^{\rm min}_i=D_j^{\rm min},
\label{didjmin}
\ee
so that
\be
\label{mainquad}
(t_i^{\rm min} \mathbf{\hat{z}} - s\mathbf{\hat{c}})^2 = (t_j^{\rm min} \mathbf{\hat{t}} + \mathbf{r} - s\mathbf{\hat{c}})^2.
\ee
Here, it is important to note that $t_i^{min}$ and $t_j^{min}$ are only the correct minimal values when $t_{i}^{\rm min} \in [-L/2,L/2]$ and $t_{j}^{\rm min} \in [-L/2,L/2]$ due to the finite length of the rods. If $t_i^{min}$ and/or $t_j^{min}$ are not $\in [-L/2,L/2]$ interactions involving the end-points of the rods arise. Overall, one has to distinguish the cases:

\begin{enumerate}
\item{{\it Line-Line} interaction: $t_{i}^{\rm min} \in [-L/2,L/2]$
    and $t_{j}^{\rm min} \in [-L/2,L/2]$ (1 case).}
\item{{\it Line-Point} interaction between the segment $i$ and an
    end-point of $j$: $t_{i}^{\rm min} \in [-L/2,L/2]$ and $t_j=\pm
    L/2$ (2 cases).}
\item{{\it Point-Line} interaction between the segment $j$ and an
    end-point of $i$: $t_{j}^{\rm min} \in [-L/2,L/2]$ and $t_i=\pm
    L/2$ (2 cases).}
\item{{\it Point-Point} interaction between the end points of $i$ and
    $j$: $t_i=\pm L/2$ and $t_j=\pm L/2$ (4 cases).}
\end{enumerate}

In the following the different Voronoi interactions are indicated by different subsscripts, e.g., $s_{\rm ll}$ for Line-Line interaction, $s_{lp}$ for Line-Point interaction, etc. Fig.~\ref{Fig_spheros_vb}b illustrates the separation of the different interactions.

\subsubsection{Line-Line interaction}

This interaction is valid if $t_i^{min}$ and $t_j^{min}$ fall inside the length of the segments. The conditions are thus:
\be
\label{tllmin}
t_{i}^{\rm min} &\in& [-L/2,L/2],\qquad t_{j}^{\rm min} \in [-L/2,L/2].
\ee
In this case $t_i^{min}$ and $t_j^{min}$ are given by Eqs.~(\ref{titj1}) and (\ref{titj2}). Substituting these expressions into Eq.~(\ref{mainquad}) yields a quadratic equation for the VB value $s=s_{\rm ll}$:
\be
\label{lleq}
\frac{s_{\rm ll}^2}{r^2} \Big[(\mathbf{\hat{c}}\mathbf{\hat{z}})^2 -(\mathbf{\hat{c}}\mathbf{\hat{t}})^2 \Big]
  + 2 \frac{s_{\rm ll}}{r} \Big[(\mathbf{\hat{c}}\mathbf{\hat{t}})(\mathbf{\hat{r}}\mathbf{\hat{t}}) - \mathbf{\hat{r}}\mathbf{\hat{c}} \Big]+ 1 - (\mathbf{\hat{r}}\mathbf{\hat{t}})^2 = 0.
\ee
The correct solution of this equation is the real and positive one. We observe that the solution $s_{\rm ll}$ scales with the separation $r$. The conditions for Eqs.~(\ref{tllmin}) to hold are
\be
-L/2\le s_{\rm ll}\mathbf{\hat{c}}\mathbf{\hat{z}}\le L/2,\quad {\rm and}\quad -L/2\le(s_{\rm ll}\mathbf{\hat{c}}-\mathbf{r})\mathbf{\hat{t}}\le L/2.
\ee
These two conditions define the separation lines for the Line-Line interaction visualized in Fig.~\ref{Fig_spheros_vb}b.

\subsubsection{Line-Point interaction}

In this case $t_j^{\rm min}$ is fixed at one of the end points of rod $j$ and $t_i^{\rm min}$ is inside the line segment of rod $i$. We set the point to the top of $\mathbf{t}_j$, indicated by a subscript $1$ and obtain:
\be
t_{i}^{\rm min} &\in& [-L/2,L/2],\qquad t_{j}^{\rm min} = L/2.
\ee
Substituting the Eq.~(\ref{titj1}) for $t_i^{min}$ and $t_{j}^{\rm min} = L/2$ into Eq.~(\ref{mainquad}) yields a quadratic equation for the VB value $s=s_{\rm lp_1}$, where the index $p_1$ refers to the top point:
\be
\label{lpeq}
  \frac{s_{\rm lp_1}^2}{r^2} (\mathbf{\hat{c}}\mathbf{\hat{z}})^2 -2\frac{s_{\rm lp_1}}{r} \Big[ (\mathbf{\hat{r}}\mathbf{\hat{c}}) 
  + \frac{L}{2r} (\mathbf{\hat{c}}\mathbf{\hat{t}}) \Big ] + \left(\frac{L}{2r}\right)^2 + \frac{L}{r} (\mathbf{\hat{r}}\mathbf{\hat{t}}) + 1=0. 
\ee
The VB generated by the bottom point $s_{\rm lp_2}$, which is defined by $t_{j}^{\rm min} = -L/2$, follows straightforwardly by setting $L\to -L$ in Eq.~(\ref{lpeq}). The conditions for the two Line-Point interactions are then (cf. Fig.~\ref{Fig_spheros_vb}b)
\be
-L/2\le s_{\rm lp_1}\mathbf{\hat{c}}\mathbf{\hat{z}}\le L/2\quad &{\rm and}&\quad (s_{\rm lp_1}\mathbf{\hat{c}}-\mathbf{r})\mathbf{\hat{t}}\ge L/2\\
-L/2 \le s_{\rm lp_2}\mathbf{\hat{c}}\mathbf{\hat{z}}\le L/2\quad &{\rm and}&\quad (s_{\rm lp_2}\mathbf{\hat{c}}-\mathbf{r})\mathbf{\hat{t}}\le -L/2.
\ee

\subsubsection{Point-Line interaction}

We can calculate this interaction in analogy to the Line-Point interaction. The conditions are:
\be
t_{i}^{\rm min} &=& L/2,\qquad t_{j}^{\rm min} \in[-L/2,L/2].
\ee
Substituting $t_{i}^{\rm min} = L/2$ for the top point and Eq.~(\ref{titj2}) into Eq.~(\ref{mainquad}) leads to
\be
\label{pleq}
&&\frac{s_{\rm p_1l}^2}{r^2} (\mathbf{\hat{c}}\mathbf{\hat{t}})^2 + 2\frac{s_{\rm p_1l}}{r} [(\mathbf{\hat{r}}\mathbf{\hat{c}}) - (\mathbf{\hat{c}}\mathbf{\hat{t}}) (\mathbf{\hat{r}}\mathbf{\hat{t}})]\nonumber\\
&&- \frac{s_{\rm p_1l}}{r}\frac{L}{r} (\mathbf{\hat{c}}\mathbf{\hat{z}})+ \left(\frac{L}{2r}\right)^2 + (\mathbf{\hat{r}}\mathbf{\hat{t}})^2- 1 = 0. 
\ee
Likewise for $s_{\rm p_2l}$. The conditions for the two point-line interactions are then
\be
s_{\rm p_1l}\mathbf{\hat{c}}\mathbf{\hat{z}}\ge L/2,\quad &{\rm and}&\quad -L/2\le (s_{\rm p_1l}\mathbf{\hat{c}}-\mathbf{r})\mathbf{\hat{t}}\le L/2\\
s_{\rm p_2l}\mathbf{\hat{c}}\mathbf{\hat{z}}\le -L/2,\quad &{\rm and}&\quad -L/2\le (s_{\rm p_2l}\mathbf{\hat{c}}-\mathbf{r})\mathbf{\hat{t}}\le L/2.
\ee

\subsubsection{Point-Point interaction}

This interaction is obtained by fixing both $t_i^{\rm min}$ and $t_j^{\rm min}$ to $L/2$ or $-L/2$. We set
\be
\label{tt}
t_i^{\rm min} = L_i/2,\qquad t_j^{\rm min} = L_j/2,
\ee
where $L_i=\pm L$ and $L_j=\pm L$ for the top and bottom points on each of the rods. The solution of Eq.~(\ref{mainquad}) with Eqs.~(\ref{tt}) is then:
\be
\label{ppsol}
  s_{pp} = r\,\frac{1 + \frac{L_j}{r} (\mathbf{\hat{r}}\mathbf{\hat{t}})}{2(\mathbf{\hat{r}}\mathbf{\hat{c}}) + \frac{L_j}{r} 
(\mathbf{\hat{c}}\mathbf{\hat{t}}) - \frac{L_i}{r}(\mathbf{\hat{c}}\mathbf{\hat{z}})}.
\ee
The different Point-Point interactions are obtained by specifying the $L_{i,j}$. The VB value $s_{\rm p_1p_1}$ due to the two top points $s_{\rm p_1p_1}$, e.g., is obtained by setting $L_i=L_j=L$. Likewise for the other point interactions. The conditions for the four different point-point Voronoi boundaries are 
\be
s_{\rm p_1p_1}\mathbf{\hat{c}}\mathbf{\hat{z}}\ge L/2,\quad &{\rm and}&\quad (s_{\rm p_1p_1}\mathbf{\hat{c}}-\mathbf{r})\mathbf{\hat{t}}\ge L/2,\\
s_{\rm p_1p_2}\mathbf{\hat{c}}\mathbf{\hat{z}}\ge L/2,\quad &{\rm and}&\quad (s_{\rm p_1p_2}\mathbf{\hat{c}}-\mathbf{r})\mathbf{\hat{t}}\le -L/2,\\
s_{\rm p_2p_1}\mathbf{\hat{c}}\mathbf{\hat{z}}\le -L/2,\quad &{\rm and}&\quad (s_{\rm p_2p_1}\mathbf{\hat{c}}-\mathbf{r})\mathbf{\hat{t}}\ge L/2,\\
s_{\rm p_2p_2}\mathbf{\hat{c}}\mathbf{\hat{z}}\le -L/2,\quad &{\rm and}&\quad (s_{\rm p_2p_2}\mathbf{\hat{c}}-\mathbf{r})\mathbf{\hat{t}}\le -L/2.
\ee
Eq.~(\ref{ppsol}) reduces to the VB between two equal spheres, Eq.~(\ref{VBsphere}), in the limit $L/r\rightarrow 0$. Note that the VBs due to the four point-point interactions are flat surfaces, while interactions involving the line segment generate curved VBs.

\subsubsection{Examples}

\begin{figure}[]
  \centering
  \includegraphics[width=0.7\textwidth]{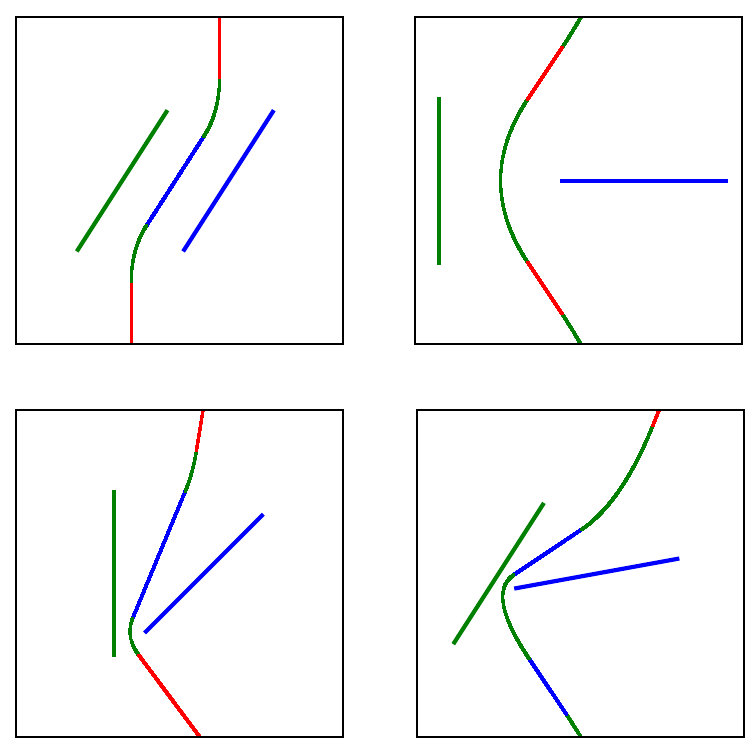}
  \caption{Examples of the VB between two rods in 2d with different configurations. Each colour
    of the boundary represents a different type of interaction. Point-Point interaction: red; Line-Point and Point-Line: green; Line-Line: blue. }
\label{2d}
\end{figure}

As an example of the algorithm, we apply it to different situations in
2 dimensions, Fig.~\ref{2d}. We consider the rod $i$ (with varying orientations) on the left
and the rod $j$ on the right. The top-left panel shows the Voronoi
boundary in different colours corresponding to different interactions:
A point-point interaction at the top of the boundary in red, then a
point-line interaction in green, then a line-line interaction in blue,
then another point-line interaction in green and so on. The other
panels are analogous.

\subsection{The contact radius of two spherocylinders}

\label{Sec_spherorad}

\begin{figure}[]
\begin{center}
\includegraphics[width=6cm]{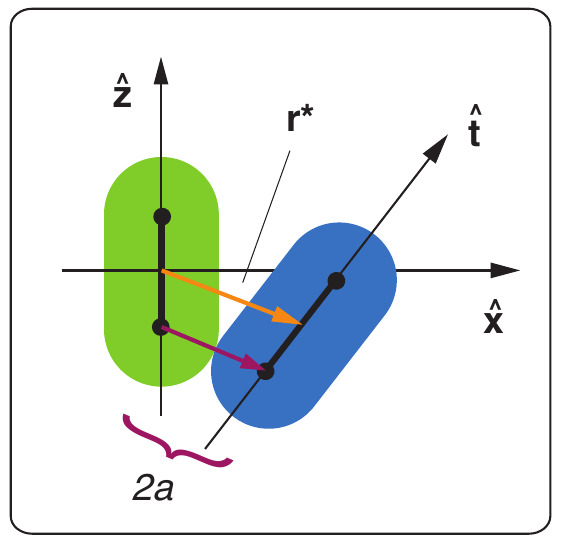} 
\caption{\label{Fig_spheros_rad} The contact radius $r^*(\mathbf{\hat{r}},\mathbf{\hat{t}})$ for two spherocylinders. Here, the contact is due to the spherical endcaps.
}
\end{center}
\end{figure}

The contact radius $r^*(\mathbf{\hat{r}},\mathbf{\hat{t}})$ is defined as the value of $r$ such that two spherocylinders of relative orientation $\mathbf{\hat{t}}$ and position $r\mathbf{\hat{r}}$ are in contact. By comparison, two spheres are in contact if $r^*=2a$, independent of $\mathbf{\hat{t}}$ and $\mathbf{\hat{r}}$. For spherocylinders we have to distinguish the possible contacts of the spherical endcaps and of the cylindrical segments (see Fig.~\ref{Fig_spheros_rad}). As before, we denote a point on rod $i$ by $t_i\mathbf{\hat{z}}$ and a point on rod $j$ by $\mathbf{r}+t_j\mathbf{\hat{t}}$. The squared distance between these two points is
 \be
 \label{Dcontact}
 D^2(\mathbf{r},\mathbf{\hat{t}},t_i,t_j)&=&(t_i\mathbf{\hat{z}}-(\mathbf{r}+t_j\mathbf{\hat{t}}))^2\nonumber\\
 &=&t_i^2+t_j^2+r^2+2r(t_j(\mathbf{\hat{r}}\mathbf{\hat{t}})-t_i(\mathbf{\hat{r}}\mathbf{\hat{z}}))-2t_it_j(\mathbf{\hat{z}}\mathbf{\hat{t}}).\nonumber\\
\ee
Two spherocylinders are in contact when the minimal $D^2$ equals the square of the diameter $(2a)^2$. We solve $\partial D^2/\partial t_i=0$ and $\partial D^2/\partial t_j=0$ in order to obtain the two minimal positions
\be
t^*_i&=&r\frac{(\mathbf{\hat{r}}\mathbf{\hat{z}})-(\mathbf{\hat{r}}\mathbf{\hat{t}})(\mathbf{\hat{z}}\mathbf{\hat{t}})}{1-(\mathbf{\hat{z}}\mathbf{\hat{t}})^2}= r A_i\\
t^*_j&=&r\frac{(\mathbf{\hat{r}}\mathbf{\hat{z}})(\mathbf{\hat{z}}\mathbf{\hat{t}})-(\mathbf{\hat{r}}\mathbf{\hat{t}})}{1-(\mathbf{\hat{z}}\mathbf{\hat{t}})^2}= r A_j,
\ee
which define $A_i$ and $A_j$. Substituting these expressions into Eq.~(\ref{Dcontact}) and solving for $r$ under the condition $D^2=4a^2$ yields the contact radius
\be
r_{\rm ll}^*(\mathbf{\hat{r}},\mathbf{\hat{t}})= \frac{2 a}{\sqrt{1+(A_i\mathbf{\hat{z}}-A_j\mathbf{\hat{t}})^2+2(A_j(\mathbf{\hat{r}}\mathbf{\hat{t}})-A_i(\mathbf{\hat{r}}\mathbf{\hat{z}}))}}.
\ee
This contact radius is only valid for $t^*_i\in[-L/2,L/2]$ and $t^*_j\in[-L/2,L/2]$, since it does not take into account the finite length of the spherocylinders. Therefore, $r_{\rm ll}^*$ is the contact between the line segments (indicated by the ${\rm ll}$ subscript as before). In complete analogy to the Voronoi interactions we need to distinguish further the Line-Point, Point-Line and Line-Line contacts in addition to the Line-Line one.

For the Line-Point contact one has to consider $t_j=\pm L/2$. In order to find the minimal $t^{\rm *lp}_i$ one thus has to solve
\be
\frac{\partial}{\partial t_i}D^2\left(\mathbf{r},\mathbf{\hat{t}},t_i,\pm \frac{L}{2}\right)=0.
\ee
Substituting this value back into $D^2$ and solving $D^2=4a^2$ for $r$ yields the two Line-Point contact radii. These are valid when $t_i^{\rm *lp}\in[-L/2,L/2]$. For the Point-Line contact one has to set $t_i=\pm L/2$, which determines the minimal $t_j^{\rm *pl}$ by the equation
\be
\frac{\partial}{\partial t_j}D^2\left(\mathbf{r},\mathbf{\hat{t}},\pm\frac{L}{2},t_j\right)=0.
\ee
Substituting this value back into $D^2$ and solving $D^2=4a^2$ for $r$ yields the two Point-Line contact radii. These are valid when $t_j^{\rm *pl}\in[-L/2,L/2]$. For the Point-Point contact one can solve directly
\be
D^2\left(\mathbf{r},\mathbf{\hat{t}},\pm\frac{L}{2},\pm\frac{L}{2}\right)=4 a^2
\ee
 for $r$, which yields four different Point-Point contact radii.

Following this procedure, we obtain 9 possible values for the contact radius $r^*(\mathbf{\hat{r}},\mathbf{\hat{t}})$, just like the 9 different VB values. Among these values correct radius is determined as the maximum of all positive and real ones.

\subsection{Calculation of the excluded volume and surface}

\label{Sec_exvols}

For completeness, we summarize here our method to calculate $V^*$ and $S^*$ for a given analytic form of the VB $s(\mathbf{r},\mathbf{\hat{t}},\mathbf{\hat{c}})$. This method has previously been implemented in Ref.~\cite{Baule13}. We define excluded Voronoi volume as $V^*=\left<\Omega-\Omega\cap V_{\rm ex}\right>_{\mathbf{\hat{t}}}$, which can be calculated as an orientational average over a volume integral:
\be
\label{exvol}
V^*(\mathbf{c})&=&\left<\int \D \mathbf{r}\,\Theta(r-r^*(\mathbf{\hat{r}},\mathbf{\hat{t}}))\Theta(c-s(\mathbf{r},\mathbf{\hat{t}},\mathbf{\hat{c}}))\Theta(s(\mathbf{r},\mathbf{\hat{t}},\mathbf{\hat{c}}))\right>_{\mathbf{\hat{t}}}.\nonumber\\
\ee
These integrals can be expressed in spherical coordinates. We denote with $\theta_r$ and $\beta_r$ the polar and azimuthal angles, respectively, of the position. The corresponding orientational angles have a subscript $t$. Eq.~(\ref{exvol}) then corresponds to the multi-dimensional integral
\begin{widetext}
\be
\label{vstar_calc}
V^*(c,\theta_{\rm c})&=&\frac{1}{2\pi}\int_0^{\pi}\D \theta_r\int_{-\pi}^\pi\D\beta_r\int_0^{\pi/2}\D \theta_t\int_{-\pi}^\pi\D\beta_t\int_{r^*(\theta_r,\beta_r,\theta_t,\beta_t)}^\infty\D r\,r^2 \sin(\theta_t)\sin(\theta_r)\nonumber\\
&&\times\Theta[c-s(r,\theta_r,\beta_r,\theta_t,\beta_t,\theta_{\rm c})]\Theta[s(r,\theta_r,\beta_r,\theta_t,\beta_t,\theta_{\rm c})].
\ee
\end{widetext}
The reduction of the full solid angle $\mathbf{\hat{c}}$ to $\theta_t$ and the integration limits of the $\theta_t$ integration take into account the symmetry of the spherocylinders. Clearly, Eq.~(\ref{vstar_calc}) is a five dimensional integral, which be calculated numerically using, e.g., a Monte-Carlo scheme for a given $\mathbf{c}$.

We define the excluded Voronoi surface as $S^*=\left<\partial V_{\rm ex}\cap\Omega\right>_{\mathbf{\hat{t}}}$, which can be expressed as an orientational average over a surface integral:
\be 
\label{exsur}
S^*(\mathbf{c})&=&\left<\left.\oint \D \mathbf{\hat{r}}\,\Theta(c-s(\mathbf{r},\mathbf{\hat{t}},\mathbf{\hat{c}}))\Theta(s(\mathbf{r},\mathbf{\hat{t}},\mathbf{\hat{c}}))\right|_{r=r^*(\mathbf{\hat{r}},\mathbf{\hat{t}})}\right>_{\mathbf{\hat{t}}},
\ee
Here, the contact radius $r^*(\mathbf{\hat{r}},\mathbf{\hat{t}})$ induces the surface element
\be
\D \mathbf{\hat{r}}=r^*\sqrt{\left(r^{*2}+\left(\frac{\partial r^*}{\partial \theta_r}\right)^2\right)\sin^2(\theta_r)+\left(\frac{\partial r^*}{\partial \beta_r}\right)^2}\D\theta_r\D\beta_r,
\ee
which recovers the usual surface element $\D \mathbf{\hat{r}}=r^{*2}\sin(\theta_r)\D\theta_r\D\beta_r$ for $r^*={\rm const}$. Eq.~(\ref{exsur}) can thus be written in terms of the four-dimensional integral
\begin{widetext}
\be
\label{sstar_calc}
S^*(c,\theta_{\rm c})&=&\frac{1}{2\pi}\int_0^{\pi}\D \theta_r\int_{-\pi}^\pi\D\beta_r\int_0^{\pi/2}\D \theta_t\int_{-\pi}^\pi\D\beta_t \sin(\theta_t)r^*\sqrt{\left(r^{*2}+\left(\frac{\partial r^*}{\partial \theta_r}\right)^2\right)\sin^2(\theta_r)+\left(\frac{\partial r^*}{\partial \beta_r}\right)^2}\nonumber\\
&&\times\Theta[c-s(r^*,\theta_r,\beta_r,\theta_t,\beta_t,\theta_{\rm c})]\Theta[s(r^*,\theta_r,\beta_r,\theta_t,\beta_t,\theta_{\rm c})], 
\ee
\end{widetext}
where $r^*=r^*(\theta_r,\beta_r,\theta_t,\beta_t)$. This expression can also be computed numerically.

\section{Discussion}

\label{Sec_discussion}

The decomposition of a shape into compositions and intersections of spheres can be generalized to more complicated shapes. In particular, the VB between any shapes that are represented by a composition or intersection of a finite number of spheres, such as trimers, tetramers, etc. with varying degree of overlap, can be calculated in a straightforward way following this method. The main challenge is to develop a concrete and efficient procedure for a given shape, i.e. a methodic protocol that is valid for a given number $n$ of overlapping spheres. The algorithm for lenses is already quite complicated. Indeed, even though the computational complexity to determine a single VB is not high, many cases have to be computed and kept in memory until the VB is identified that specifies the correct interaction. There are additional technical difficulties for the separation of the interactions. For lens-shaped particles, e.g., polynomials of order 6 have to be solved numerically.

Shapes of particular recent interest in theoretical materials science are polyhedra, which can self-assemble into structures much more complicated than spheres and have thus great potential for the assembly of new functional materials \cite{Escobedo11,Damasceno12}. In order to apply our framework to polyhedra, one can consider a decomposition of the shape into a dense union of spheres with continuously varying radii. This kind of decomposition (also called filling) can be optimized following certain design principles \cite{Phillips12}. Even though our method to determine the VB is still algorithmically well-defined for such a dense filling, it is difficult to implement in practice. A simpler approach would be to approximate a polyhdra as an intersection of a small finite number of spheres, similar to the approximation of an ellipsoid by a lens-shaped particle discussed here. A cube, e.g., can be approximated as the intersection of six spheres \cite{Baule13}, such that the VB can be calculated following our method.

Even if an analytical expression for the VB of a particular shape can be obtained, the calculation of the excluded volume and surface are still computationally costly due to the high dimensional integrals. The calculation of the Jacobian provides additional difficulties. A direct numerical computation of the VB might provide an alternative approach, which can be optimized for speed using, e.g., graphics hardware \cite{Hoff99}.

The code used to generate the VB for lens-shaped particles and spherocylinders is available on www.jamlab.org.

\begin{acknowledgements}

We gratefully acknowledge funding by NSF-CMMT and DOE Office of Basic Energy Sciences, Chemical
Sciences, Geosciences, and Biosciences Division.

\end{acknowledgements}


\end{document}